\documentclass[%
 aip,
 amsmath,amssymb,
reprint,%
]{revtex4-1}

\usepackage[pdftex]{graphicx}
\usepackage{dcolumn}
\usepackage{bm}
\usepackage[version=3]{mhchem}

\usepackage[utf8]{inputenc}
\usepackage[T1]{fontenc}
\usepackage{mathptmx}
\usepackage{etoolbox}

\makeatletter
\def\@email#1#2{%
 \endgroup
 \patchcmd{\titleblock@produce}
  {\frontmatter@RRAPformat}
  {\frontmatter@RRAPformat{\produce@RRAP{*#1\href{mailto:#2}{#2}}}\frontmatter@RRAPformat}
  {}{}
}%
\makeatother
\begin{document}

\preprint{AIP/123-QED}

\title{Beyond ab initio reaction simulator: an application to GaN metalorganic vapor phase epitaxy}

\author{A. Kusaba}
 \email{kusaba@riam.kyushu-u.ac.jp}
\affiliation{ 
Research Institute for Applied Mechanics, Kyushu University, Kasuga, Fukuoka 816-8580, Japan
}

\author{S. Nitta}
\author{K. Shiraishi}
\affiliation{
Institute of Materials and Systems for Sustainability, Nagoya University, Chikusa-ku, Nagoya 464-8601, Japan
}

\author{T. Kuboyama}
\affiliation{
Computer Centre, Gakushuin University, Toshima-ku, Tokyo 171-8588, Japan
}

\author{Y. Kangawa}
\affiliation{ 
Research Institute for Applied Mechanics, Kyushu University, Kasuga, Fukuoka 816-8580, Japan
}

\date{\today}

\begin{abstract}
To develop a quantitative reaction simulator, data assimilation was performed using high-resolution time-of-flight mass spectrometry (TOF-MS) data applied to GaN metalorganic vapor phase epitaxy system. Incorporating ab initio knowledge into the optimization successfully reproduces not only the concentration of \ce{CH_4} (an impurity precursor) as an objective variable but also known reaction pathways. The simulation results show significant production of \ce{GaH3}, a precursor of GaN, which has been difficult to detect in TOF-MS experiments. Our proposed approach is expected to be applicable to other applied physics fields that require quantitative prediction that goes beyond ab initio reaction rates.
\end{abstract}

\maketitle


Crystal growth technology of III-nitride semiconductors has been actively developed to fabricate optical and electronic devices \cite{amano1986metalorganic,palacios2005high,otake2008vertical,kachi2014recent,oka20151,sato2020room}. Reactor simulators that solve for heat and mass transfer, chemical reactions, and phase transitions have played a role in these developments \cite{ohkawa2001growth,karpov2003advances,Zuo2012P46,tseng2015transport,li2018study,niedzielski2021numerical}. The reaction model and its kinetic parameters are key components of reactor simulators, especially for the reaction system for chemical vapor deposition. It is generally quite difficult to determine the complete kinetics of complex gas-phase reaction networks using experiments alone. Thus, such reaction systems have been studied by ab initio calculations, e.g., density functional theory (DFT) and transition state theory (TST) \cite{nakamura2000quantum,makino2000quantum,mazumder2001importance,sengupta2003does,cavallotti2004accelerated,hirako2005modeling,moscatelli2007theoretical,ravasio2015analysis,sakakibara2021theoretical}. Increased semiconductor device performance has recently necessitated the use of simulators for quantitative process optimization that goes beyond obtaining a qualitative understanding \cite{dropka2021development,schimmel2021numerical,ghritli2021estimation,kusaba2022exploration}. However, the predictive performance of ab initio reaction rates is often inadequate for such quantitative requirements.

The aim of this study is to realize quantitative simulation by data assimilation for a reaction model with rate constant parameters for the GaN metalorganic vapor phase epitaxy (MOVPE) system. A complex reaction network generally involves many parameters, and experimental measurements are costly and provide limited data. Under these circumstances, two measures are taken to ensure that the solution is not indefinite or an overfit. First, a reduced reaction model is adopted that excludes many radical reactions and consists of fewer reactions than conventional models \cite{sakakibara2021theoretical}. This reduced model has been demonstrated to reproduce the latest experimental fact on the reaction pathway that the removal of the methyl group in trimethylgallium (TMG) decomposition is caused by reactions with \ce{NH3} \cite{ye2021cgct8}. Second, the optimization is performed with the strategy that not only focuses on minimizing simulation errors, but also maintains the theoretical basis from ab initio calculations. In addition, the reaction pathway becomes a fundamental consideration in selecting a solution. This approach is used to develop a reliable and quantitative reaction simulator in this study.

Experimental data were obtained by high-resolution time-of-flight mass spectrometry (TOF-MS) \cite{ye2021cgct8,ye2019ammonia,ye2020analysis} as the detection intensity of the individual molecules in MOVPE environments. This state-of-the-art mass spectrometric technique can distinguish between \ce{NH2} and \ce{CH4}, which is not possible using conventional quadrupole mass spectrometry (QMS) \cite{nagamatsu2017decomposition}. Indeed, this study relies on the \ce{CH4} concentration data converted from its intensity data. Details of the data conversion are presented in Appendix A. Experimental measurements were performed dozens of times at different temperatures, but only data with kinetic information were used for data assimilation.

The input‒output relationship of the simulator is concisely expressed as
\begin{eqnarray}
\hat{c}(x) = f(T_{set};k).
\end{eqnarray}
Here, the output $\hat{c}(x)$ is the concentration distribution, the input $T_{set}$ is the heater temperature (an experimental condition), the simulation parameter $k$ is the reaction rate constant, and the simulator $f$ solves the ordinary differential equations of the reaction kinetics for our reduced reaction model shown in Table I. The reaction rate constant is determined using DFT: B3LYP/6-311G(d,p) and TST \cite{sakakibara2021theoretical}. The modified Arrhenius format given below is used to model the temperature dependence of the ab initio rate constant and subsequently tuned to better reproduce the experimental data:
\begin{eqnarray}
k = (q_AA)T\exp\left(-\frac{(q_EE_{act})}{k_bT}\right).
\end{eqnarray}
Here, the parameters $E_{act}$ and $A$ are the activation energy and preexponential factor obtained from ab initio calculations, and $q_E$ and $q_A$ are tuning coefficients for these parameters. Note that $E_{act}$ is not equal to the simple DFT activation energy because of the effects of TST. More details on the simulator can be found in Appendix B.

A multiobjective optimization was performed to determine the $k$ (i.e., $q_E$ and $q_A$) that maintains the theoretical basis obtained from ab initio calculations and reproduces the experimental data well. The ab initio ratios of $E_{act}$ were used as the theoretical basis. The concentration of \ce{CH4} (an impurity precursor) was used as a measure of the simulation performance because \ce{CH4} is a stable molecule and thus, the corresponding detection data are reliable and can be converted to concentration data by a reasonable scheme. Our optimization problem is expressed as follows:
\begin{equation}
minimize
\left\{ \,
\begin{aligned}
& \sum_i \left( \frac{ \hat{c}_{\ce{CH4}}(x_d;{T_{set}}_i,k) - c_{\ce{CH4}}({T_{set}}_i) }{ c_{\ce{CH4}}({T_{set}}_i) } \right)^2 \\
& \frac{1}{N} \sum_j \left( {q_E}_j - \overline{{q_E}_j} \right)^2
\end{aligned}.
\right.
\end{equation}
The first objective function is based on the relative error in the \ce{CH4} concentration between the experimental ($c_{\ce{CH4}}$) and simulation ($\hat{c}_{\ce{CH4}}$) results at the detection position $x_d$, where $i$ is the index of the experimental conditions. In this case, only $T_{set}$ was varied. The second objective function is the variance of $q_E$, where $j$ is the reaction index and $N$ is the total number of considered reactions. The smaller the variance is (i.e., the closer the coefficients ${q_E}_j$ are to each other), the closer the ratios of the modified activation energies (${q_E}_j{E_{act}}_j/{q_E}_k{E_{act}}_k$) are to the ab initio ratios (${E_{act}}_j/{E_{act}}_k$). Multiobjective optimization determines the best trade-off solutions (i.e., Pareto solutions) rather than just one optimal solution. Evolutionary algorithms are widely used for finding Pareto solutions to multiobjective optimization problems. A well-known algorithm called the fast elitist nondominated sorting genetic algorithm (NSGA-II) was employed in this study \cite{deb2002fast}.

Although the parameters are softly constrained by the second objective function, the reaction model consisting of 29 reactions still has 58 parameters. To avoid difficulties in interpreting the tuning results, hard constraints are also imposed by requiring $q_E$ to be the same for each reaction group (Groups G1–G5, R7, and R14 in Table I). That is, the chemical equations belonging to the same group have the same reaction partners (\ce{H2} or \ce{NH3}) and byproducts (\ce{CH4}, \ce{NH3}, or \ce{H2}) and differ only in the substituents (\ce{-CH3}, \ce{-NH2}, or \ce{-H}) not involved in the reaction. Thus, these equations are expected to have the same degree of error in the ab initio $E_{act}$. In addition, as the preexponential factor does not have as large an influence as the activation energy, the same $q_A$ was assumed for all reactions for simplicity.

Figure~1(a) shows the Pareto solutions obtained from 30,000 function evaluations in the optimization. Here, a marker represents a set of tuning coefficients, i.e., $q_E$ for each reaction group and $q_A$. These approximate Pareto solutions have the smallest variance of $q_E$ among the many possible sets of coefficients resulting in the same error. The solutions correspond to a trade-off between the two objective functions. That is, the larger the variance of $q_E$ is, the lower the root mean squared percentage error (RMSPE) of \ce{CH4} concentration is. However, even with a solution that emphasizes lowering the variance of $q_E$, the RMSPE is only less than 25\%. In Fig.~1(b), the horizontal axis is $q_E$ and the vertical axis is the RMSPE. A marker in Fig.~1(a) corresponds to 7 markers aligned horizontally in Fig.~1(b) at the same RMSPE level. A solution with an RMSPE of approximately 25\% has an almost identical $q_E$ of approximately 0.85 for each reaction group, that is, this solution maintains the ab initio ratios of $E_{act}$. The solutions that emphasize reducing the RMSPE have different $q_E$ values for each reaction group. For example, the solution with an RMSPE of 15.0\% presented in Fig.~1(a) has a set of $q_E$s at a 15.0\% RMSPE level in Fig.~1(b) (i.e., 0.795, 0.885, 0.846, 0.849, 0.848, 0.850 and 0.846 for G1, G2, G3, G4, G5, R7 and R14, respectively). To reduce the RMSPE, the $q_E$ for Group G1 needs to be relatively decreased, and the $q_E$ for Group G2 needs to be relatively increased. Note that $q_EE_{act}$ for all reactions is an absolute value that has been decreased from the ab initio $E_{act}$ to obtain these Pareto solutions. In addition, $q_A$ exhibited fluctuations within 20\% corresponding to discontinuous changes in $q_E$.

\begin{figure}
\includegraphics[width=8.5cm]{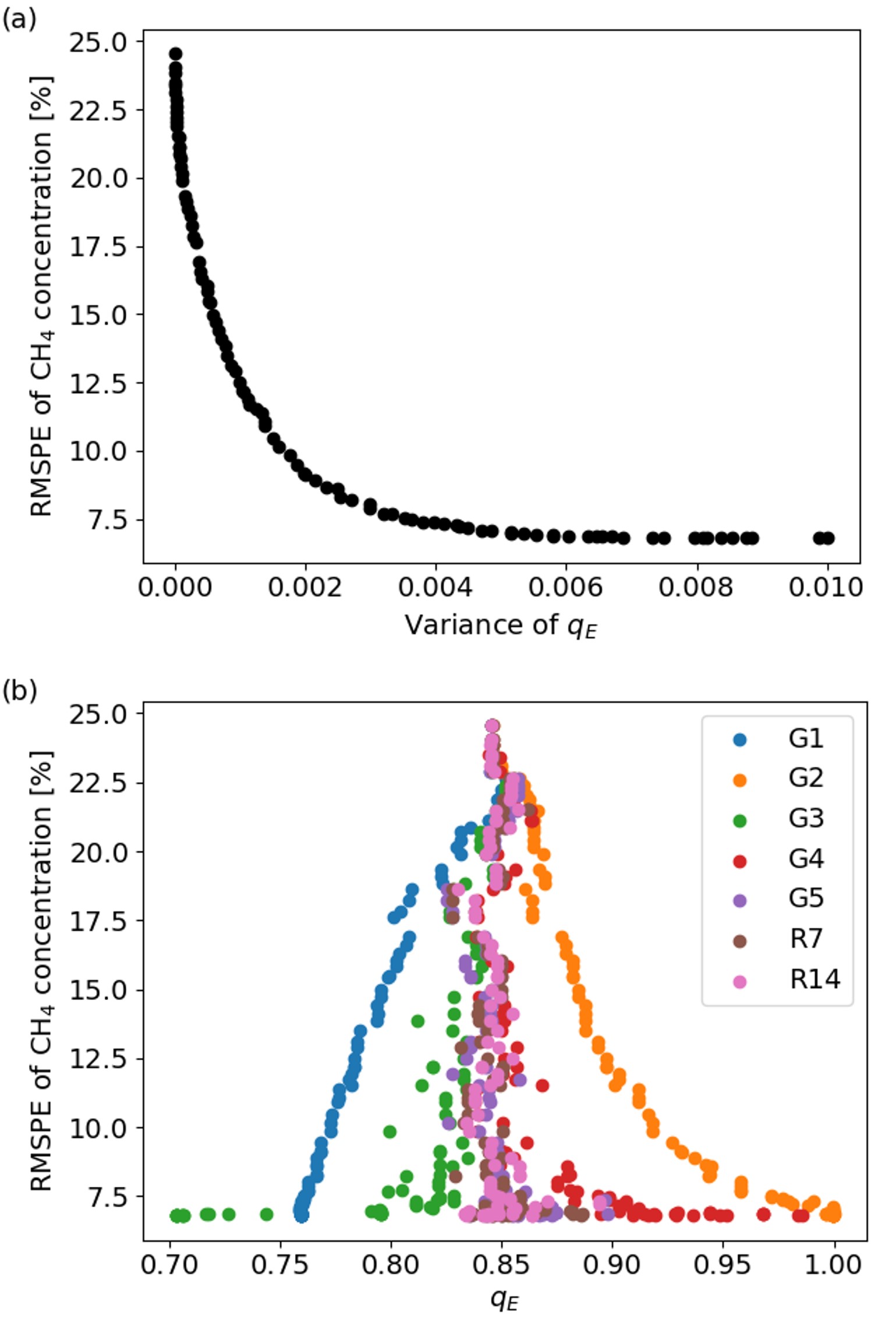}
\caption{(a) Pareto solutions for two objective functions obtained by NSGA-II. The first objective function is expressed in terms of the root mean squared percentage error (RMSPE). (b) The RMSPE dependence of the $q_E$ component for the different reaction groups of the solutions shown in (a).}
\end{figure}

Figure~2 is a comparison of the simulation performance between the parameter sets tuned by the optimization and the original, untuned parameter set. The difference in the observed \ce{CH4} concentration corresponds to the difference in the experimental conditions $T_{set}$. Simulations were performed at each $T_{set}$ using a number of tuned parameter sets and the original parameter set. The results obtained using the tuned parameter sets are plotted in different colors depending on the variance of $q_E$. The parameter sets with the largest and smallest $q_E$ variance among the parameter sets shown in Fig.~1 are hereafter referred to as the experimental and ab initio limits, respectively. That is, the experimental and ab initio limits are the solutions with the lowest error in the \ce{CH4} concentration and that best preserve the ab initio $E_{act}$ ratios, respectively. The closer the markers in Fig.~2 are to the proportionality line, the better the simulation performance is. Using the original parameter set resulted in a considerable underestimate of the \ce{CH4} concentration. Lowering $E_{act}$ by using the tuning factors shown in Fig.~1(b) promoted the TMG decomposition reactions (Groups G1, G2, R7, and R14) and increased \ce{CH4} production up to near the observed values. It is remarkable that sufficiently quantitative simulations have been performed that represent a significant improvement over the original, even in the case of the ab initio limit.

\begin{figure}
\includegraphics[width=9cm]{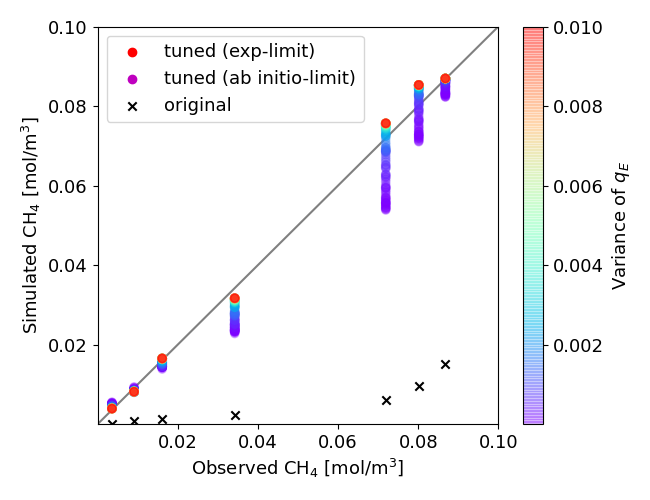}
\caption{Simulated \ce{CH4} concentration values obtained using the original parameter set and tuned parameter sets versus the experimental data. The marker color for the tuned parameter sets changes with the variance of $q_E$.}
\end{figure}

Next, we discuss how a solution should be selected among the Pareto solutions. It is nontrivial to realize a proper balance between the two trade-off objective functions. Thus, we consider this problem from a different point of view than that considered for the optimization. Figure~3 shows the reaction rates for (a) R1 and (b) R8 at axial positions in the reactor (i.e., $k_{\rm{R1}}[\rm{TMG}][\ce{H2}]$ and $k_{\rm{R8}}[\rm{TMG}][\ce{NH3}]$). Here, the heater position ranges from 0 to 30 [cm] and the TOF-MS detection position is 20 [cm]. It can be seen that Reaction R8 dominates over Reaction R1 in both the original and ab initio limits. Reaction R1 becomes dominant in the experimental limit. On the other hand, it was reported that in a TOF-MS observation using a heavy hydrogen \ce{D2} supply instead of \ce{H2}, \ce{CH4} or \ce{CH3D} was dominantly detected for the condition with or without \ce{NH3}, respectively \cite{ye2021cgct8}. This experimental result indicates that the main pathway involves TMG decomposition reactions with \ce{NH3} (i.e., Group G2) and not reactions with \ce{H2} (i.e., Group G1). In addition, this reaction pathway was qualitatively supported by an ab initio study \cite{sakakibara2021theoretical}. Thus, the ab initio limit solution should be selected to be consistent with the experimentally and theoretically supported reaction pathways. Note that the knowledge of the reaction pathway was not included in the optimization, although the results between the ab initio and experimental limits (indicated by color gradations) are trade-offs in terms of the two objective functions.

\begin{figure}
\includegraphics[width=9cm]{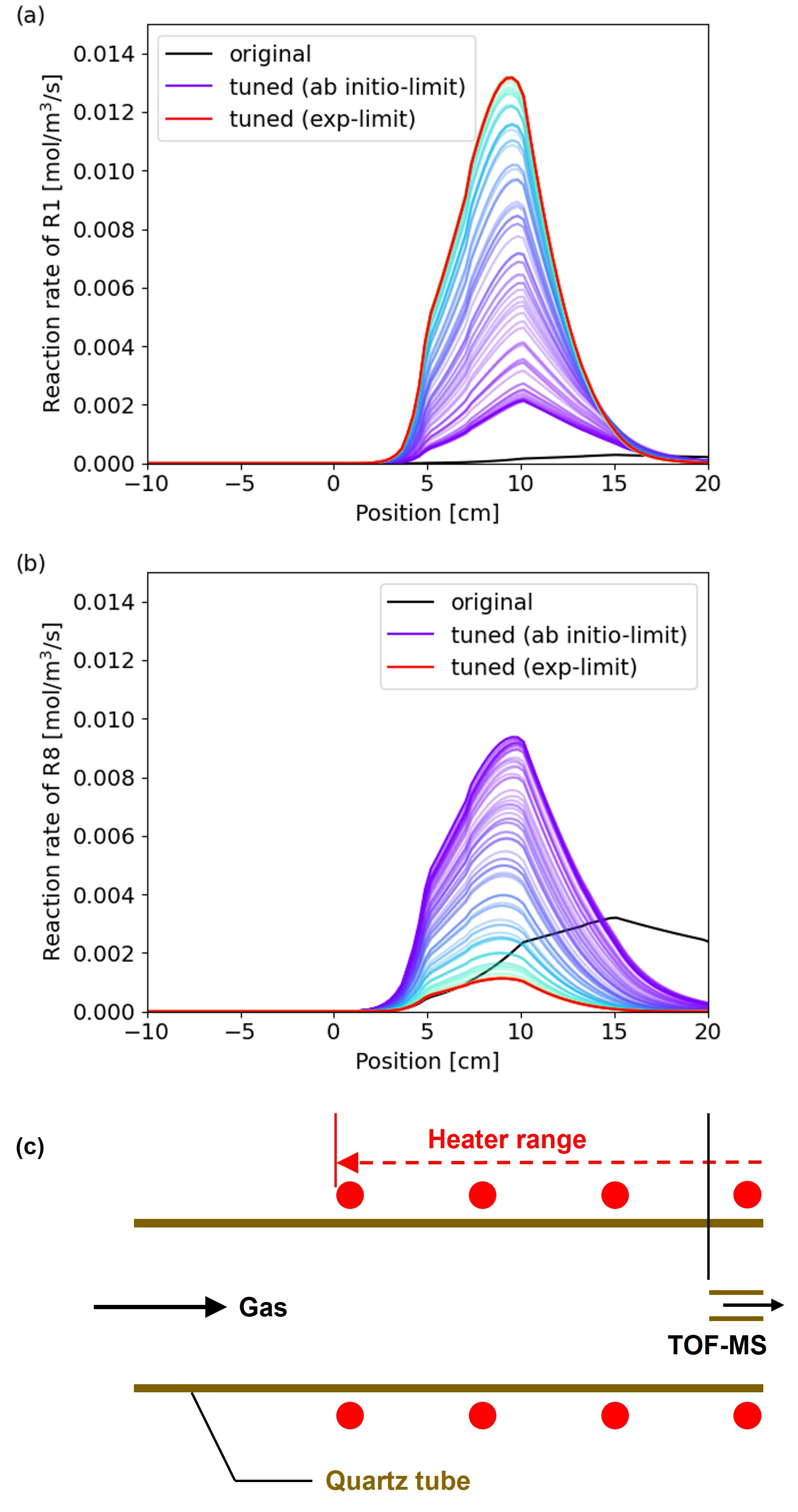}
\caption{Rection rate distributions determined using the original and tuned parameter sets for (a) Reaction R1: \ce{Ga(CH3)3 + H2 -> Ga(CH3)2H + CH4} and (b) Reaction R8: \ce{Ga(CH3)3 + NH3 -> Ga(CH3)2NH2 + CH4}. The plot color for the tuned parameter sets corresponds to the color bar in Fig.~2. (c) Schematic of a reactor equipped with TOF-MS.}
\end{figure}

As a result of our quantitative reaction simulations, it is worth focusing on the concentration of \ce{GaH3} because it is considered a precursor of the GaN film. Figure~4(a) shows the concentration distribution of \ce{GaH3} at $T_{set}=600$ [℃]. It can be seen that \ce{GaH3} is produced in a significant yield under TOF-MS conditions and is the main precursor considering that the TMG supply is 0.029 mol/m$^3$. Note that Ga-related products are difficult to detect in TOF-MS experiments under high-temperature conditions, for some reasons \cite{ye2020analysis}. That is, the complementary use of TOF-MS observations and data-assimilated reaction simulations can enable a more detailed study of gaseous compositions than previously possible. On the other hand, other precursor candidates GaH, \ce{GaCH3}, and \ce{GaNH2} were little produced. Nevertheless, even the rates of reactions producing these precursors are considered meaningful in the ab initio limit.

\begin{figure}
\includegraphics[width=8cm]{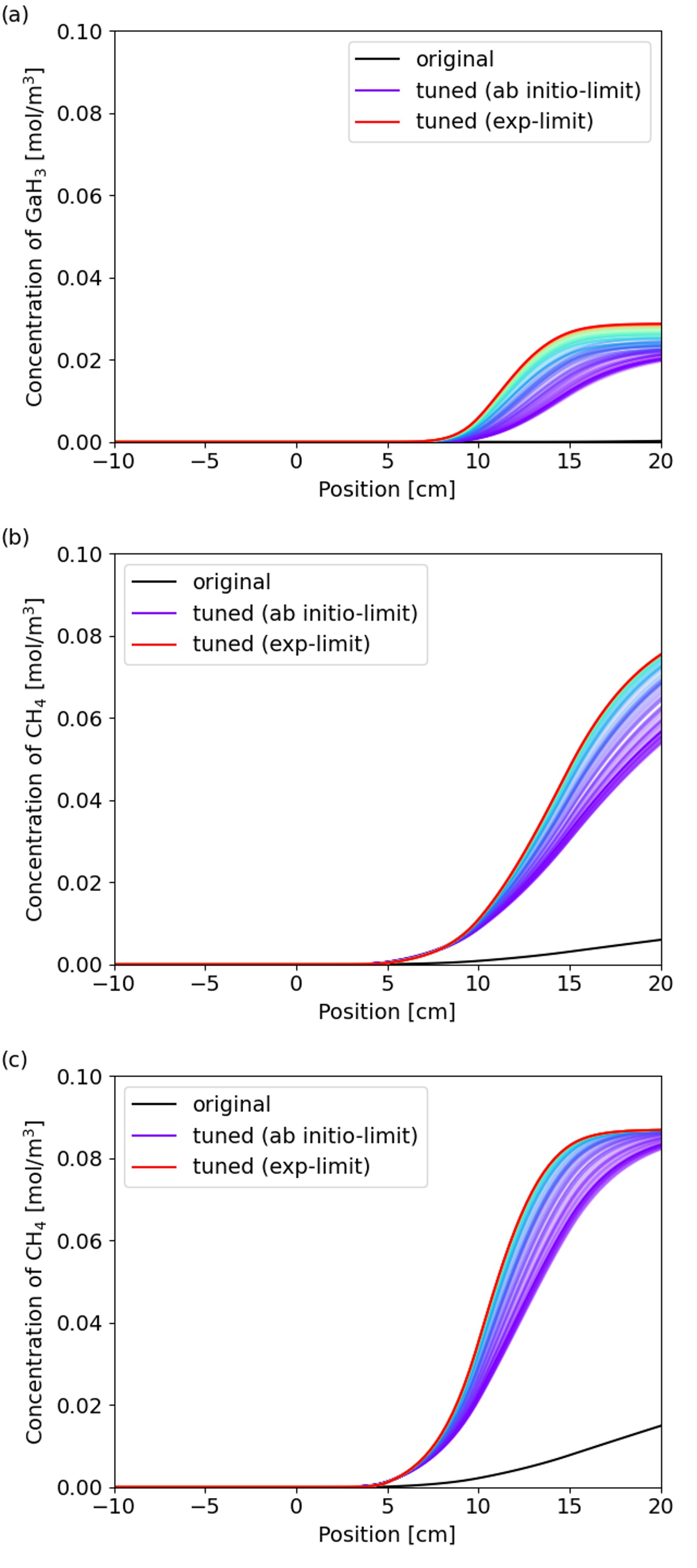}
\caption{Concentration distribution determined using the original and tuned parameter sets for (a) \ce{GaH3} at $T_{set}=600$ [℃], (b) \ce{CH4} at $T_{set}=550$ [℃] and (c) \ce{CH4} at $T_{set}=600$ [℃]. The plot color for the tuned parameter sets corresponds to the color bar in Fig.~2.}
\end{figure}

An experimental protocol for better tuning can be discussed in terms of the deviation of the simulated value. Figure~2 also shows that the deviation of the \ce{CH4} concentration simulated using the Pareto solutions differs by the observed value resulting from $T_{set}$. For example, the deviation of the simulation for the observation of 0.072 mol/m$^3$ at $T_{set}=550$ [℃] is larger than that for the observation of 0.087 mol/m$^3$ at $T_{set}=600$ [℃]. Figures~4(b) and 4(c) show the simulated \ce{CH4} concentration distributions at $T_{set}=550$ and $600$ [℃], respectively. The values at the detection position 20 [cm] shown in Figs.~4(b) and 4(c) correspond to the data shown in Fig.~2. Immediately after the decomposition reactions begin and when the reactions are almost complete, the deviation of the simulation is small, then the observation data are relatively uninformative as shown in Fig.~4(c). In the transient or kinetic domain, the deviation of the simulation is large, and then the observation data are relatively informative as shown in Fig.~4(b). For more informative observation data, as $T_{set}$ rises, the detection position should be adjusted to the inlet side or the flow velocity should be adjusted faster. Using this protocol to collect more experimental data would enable more reliable rate constants to be determined.

In conclusion, data assimilation of the gas-phase reaction rate in GaN MOVPE was performed by multiobjective optimization. Simple optimization to reproduce the experimental \ce{CH4} concentration did not yield reaction rates that reproduce known major reaction pathways. An approach was proposed to resolve this difficulty in which rate constants are tuned by imposing appropriate constraints based on ab initio calculations. The developed quantitative reaction simulator is a powerful tool for GaN MOVPE research. In addition, the proposed tuning approach is expected to be applicable to other applied physics fields that require quantitative prediction that goes beyond ab initio reaction rates.

\begin{table*}
\caption{List of reactions in the reduced gas phase reaction model of GaN MOVPE. The parameters in $k=AT\exp(-E_{act}⁄k_b T)$ are based on ab initio calculations \cite{sakakibara2021theoretical}. Units of $A$ and $E_{act}⁄k_b$  are m$^3$/mol/s/K and K. X, Y = \ce{-CH3}, \ce{-NH2} or \ce{-H}.}
\begin{ruledtabular}
\begin{tabular}{lccc}
ID & Reaction & $A$ & $E_{act}/k_b$\\
\hline
\multicolumn{4}{c}{G1: \ce{GaXYCH3 + H2 -> GaXYH + CH4}}     \\
R1 & \ce{Ga(CH3)3 + H2 -> Ga(CH3)2H + CH4} & 4.14 $\times10^2$ & 17506 \\
R2 & \ce{Ga(CH3)2H + H2 -> GaCH3H2 + CH4} & 5.06 $\times10^2$ & 17203 \\
R3 & \ce{GaCH3H2 + H2 -> GaH3 + CH4} & 4.98 $\times10^2$ & 16776 \\
R4 & \ce{Ga(CH3)2NH2 + H2 -> GaCH3HNH2 + CH4} & 1.83 $\times10^3$ & 22753 \\
R5 & \ce{GaCH3HNH2 + H2 -> GaH2NH2 + CH4} & 1.93 $\times10^3$ & 22739 \\
R6 & \ce{GaCH3(NH2)2 + H2 -> GaH(NH2)2 + CH4} & 3.97 $\times10^2$ & 26051 \\
\multicolumn{4}{c}{X, Y = none in G1}     \\
R7 & \ce{GaCH3 + H2 -> GaH + CH4} & 3.07 $\times10^3$ & 13466 \\
\multicolumn{4}{c}{G2: \ce{GaXYCH3 + NH3 -> GaXYNH2 + CH4}}     \\
R8 & \ce{Ga(CH3)3 + NH3 -> Ga(CH3)2NH2 + CH4} & 3.22 $\times10^1$ & 11096 \\
R9 & \ce{Ga(CH3)2H + NH3 -> GaCH3HNH2 + CH4} & 3.39 $\times10^1$ & 10913 \\
R10 & \ce{GaCH3H2 + NH3 -> GaH2NH2 + CH4} & 2.88 $\times10^1$ & 10651 \\
R11 & \ce{Ga(CH3)2NH2 + NH3 -> GaCH3(NH2)2 + CH4} & 7.03 $\times10^1$ & 14575 \\
R12 & \ce{GaCH3HNH2 + NH3 -> GaH(NH2)2 + CH4} & 1.86 $\times10^2$ & 14563 \\
R13 & \ce{GaCH3(NH2)2 + NH3 -> Ga(NH2)3 + CH4} & 3.07 $\times10^1$ & 16575 \\
\multicolumn{4}{c}{X, Y = none in G2}     \\
R14 & \ce{GaCH3 + NH3 -> GaNH2 + CH4} & 3.88 $\times10^2$ & 10226 \\
\multicolumn{4}{c}{G3: \ce{GaXYNH2 + H2 -> GaXYH + NH3}}     \\
R15 & \ce{Ga(CH3)2NH2 + H2 -> Ga(CH3)2H + NH3} & 4.55 $\times10^2$ & 12993 \\
R16 & \ce{GaCH3HNH2 + H2 -> GaCH3H2 + NH3} & 5.26 $\times10^2$ & 12928 \\
R17 & \ce{GaH2NH2 + H2 -> GaH3 + NH3} & 4.53 $\times10^2$ & 12949 \\
R18 & \ce{GaCH3(NH2)2 + H2 -> GaCH3HNH2 + NH3} & 1.64 $\times10^2$ & 13853 \\
R19 & \ce{GaH(NH2)2 + H2 -> GaH2NH2 + NH3} & 1.42 $\times10^2$ & 14012 \\
R20 & \ce{Ga(NH2)3 + H2 -> GaH(NH2)2 + NH3} & 1.54 $\times10^2$ & 14912 \\
\multicolumn{4}{c}{G4: \ce{GaXYH + NH3 -> GaXYNH2 + H2}}     \\
R21 & \ce{Ga(CH3)2H + NH3 -> Ga(CH3)2NH2 + H2} & 4.35 $\times10^1$ & 9707.4 \\
R22 & \ce{GaCH3H2 + NH3 -> GaCH3HNH2 + H2} & 4.44 $\times10^1$ & 9510.6 \\
R23 & \ce{GaH3 + NH3 -> GaH2NH2 + H2} & 4.94 $\times10^1$ & 9373.6 \\
R24 & \ce{GaCH3HNH2 + NH3 -> GaCH3(NH2)2 + H2} & 9.71 $\times10^1$ & 12955 \\
R25 & \ce{GaH2NH2 + NH3 -> GaH(NH2)2 + H2} & 8.30 $\times10^1$ & 13050 \\
R26 & \ce{GaH(NH2)2 + NH3 -> Ga(NH2)3 + H2} & 5.24 $\times10^1$ & 15155 \\
\multicolumn{4}{c}{G5: \ce{GaXH2 + H2 -> GaX + 2H2}}     \\
R27 & \ce{GaH3 + H2 -> GaH + 2H2} & 1.52 $\times10^3$ & 26439 \\
R28 & \ce{GaCH3H2 + H2 -> GaCH3 + 2H2} & 1.42 $\times10^3$ & 27154 \\
R29 & \ce{GaH2NH2 + H2 -> GaNH2 + 2H2} & 3.48 $\times10^4$ & 27600 \\
\end{tabular}
\end{ruledtabular}
\end{table*}

\begin{acknowledgments}
The author (A.K.) wishes to acknowledge Dr. Z. Ye and Prof. K. Yumimoto for their valuable advice.
This study was supported by JST ACT-X (Grant No. JPMJAX20A7), JSPS KAKENHI (Grant No. JP20K15181), and MEXT “Program for Promoting Researches on the Supercomputer Fugaku” (Grant No. JPMXP1020200205).
\end{acknowledgments}

\appendix
\section{Conversion of TOF-MS Intensity Data to the \ce{CH4} Concentration}
The \ce{CH4} intensity resulting from \ce{Ga(CH3)3} decomposition under a \ce{H2} atmosphere in the absence of \ce{NH3} (see Fig. 6(c) of Ref. 28) depends on the temperature as follows: the intensity is constant at $T_{set}\leq400$ [℃], rapidly increases at $425\leq T_{set}\leq525$ [℃], and then becomes constant again at $T_{set}\geq550$ [℃]. These data can be interpreted as follows: \ce{Ga(CH3)3} decomposition reactions are not initiated in the low-temperature domain ($T_{set}\leq400$ [℃]), proceed in the kinetic domain ($425\leq T_{set}\leq525$ [℃]), and are completed in the equilibrium domain ($T_{set}\geq550$ [℃]). Thus, the \ce{CH4} concentration is considered zero in the low-temperature domain and to be three times the \ce{Ga(CH3)3} supply in the equilibrium domain because the full decomposition reaction is \ce{Ga(CH3)3 + 3H2 -> GaH3 + 3CH4}.  The \ce{CH4} intensity exhibits a similar temperature dependence for the case of a \ce{H2} atmosphere containing \ce{NH3}, which corresponds to the system on which we performed data assimilation (note that the \ce{CH4} intensity decreases again at higher temperatures to react with \ce{NH3}, but this behavior is out of the scope of our reduced reaction model)$^{26}$. If we assume that the intensity and concentration are proportional to each other in the kinetic domain, the data conversion can be expressed as follows:
\begin{eqnarray}
c_{\ce{CH4}}(T_{set}) = 3c^0_{\ce{TMG}}\frac{I(T_{set})-I_{min}}{I_{max}-I_{min}}.
\end{eqnarray}
Here, $c_{\ce{CH4}}$  is the converted \ce{CH4} concentration, $c^0_{\ce{TMG}}$ is the concentration of supplied \ce{Ga(CH3)3}, $I$ is the \ce{CH4} intensity, $I_{max}$ corresponds to the completely decomposed situation, and $I_{min}$ corresponds to the completely undecomposed situation.

\section{Reaction Simulator}
The ordinary differential equations for the reaction kinetics are expressed as follows:
\begin{eqnarray}
\frac{dc_i(t)}{dt} = f(c_1(t), c_2(t), \cdots, k_{\rm{R1}}(T), k_{\rm{R2}}(T), \cdots).
\end{eqnarray}
Here, $c_i$ is the concentration of molecule $i$, $k_j$ is the rate constant for reaction $j$=R1,R2,$\cdots$ in Table I, $t$ is the time and $T$ is the temperature. The temperature distribution $T(x)$ is needed to solve Eq. (B1). The experimental temperature distribution obtained from thermocouple measurements along the central axis of a cylindrical reactor under a \ce{H2} atmosphere was used in this study (see Fig. 2 of Ref. 28). As the temperature distribution data were obtained in 5-cm increments with respect to the position and in 100 ℃ increments with respect to $T_{set}$, linear interpolation was performed for each. Next, the flow velocity was approximately calculated from the supply flow rate and the cross-sectional area of the cylinder. If the flow is captured by the Lagrangian specification, the position is a function of time, i.e., $x(t)$. Thus, $T$ in Eq. (B1) is expressed as follows:
\begin{eqnarray}
T=T(x(t))
\end{eqnarray}
This implies that the chemical reactions are solved within a reference frame that moves with the flow along the central axis. Note that the intensity detection position also lies along the central axis.

%

\end{document}